\documentclass{ws-ijmpa}
\usepackage[super,compress]{cite}
\usepackage{graphicx}
\usepackage{url}
\usepackage{hyperref}
\usepackage[utf8]{inputenc}
\begin{document}

\newcommand{\BE}{\begin{equation}}
\newcommand{\EE}{\end{equation}}
\newcommand{\half}{{\scriptstyle{\frac{1}{2}}}}

\bibliographystyle{ws-ijmpa}

\markboth{M. Consoli, L.Cosmai}{Experimental signals for a second resonance of the Higgs field}

%
\catchline{}{}{}{}{}
%

\title{Experimental signals for a second resonance of the Higgs field}

\author{Maurizio Consoli}

\address{INFN - Sezione di Catania,  I-95129 Catania, Italy \\
maurizio.consoli@ct.infn.it}

\author{Leonardo Cosmai}

\address{INFN - Sezione di Bari, I-70126 Bari, Italy\\
leonardo.cosmai@ba.infn.it}

\maketitle

\begin{history}
\received{Day Month Year}
\revised{Day Month Year}
\end{history}

\maketitle

\begin{abstract}
\noindent In the region of invariant mass 620$\div$740 GeV, we have
analyzed the ATLAS sample of 4-lepton events that could indicate a
new scalar resonance produced mainly via gluon-gluon fusion. These
data suggest the existence of a new heavy state $H$ whose mass
$660\div 680$ GeV would fit well with the theoretical range $M_H =
690 \pm 10 ~({\rm stat}) \pm 20 ~({\rm sys})~ {\rm GeV}$ for the
hypothetical second resonance of the Higgs field that has been
recently proposed and which would couple to longitudinal W's with
the same typical strength of the low-mass state at $125$ GeV. Since
the total width $\Gamma_H$ is very poorly determined, to sharpen the
analysis of the precious ATLAS data, we have considered a particular
correlation between resonating peak cross section $\sigma_R(pp\to H
\to 4l)$ and the ratio $\gamma_H=\Gamma_H/M_H$. This correlation
should be nearly insensitive to the precise value of $\Gamma_H$ and
mainly determined by the lower mass $m_h=$ 125 GeV. Equivalently, if
this correlation holds true, one could also fit $m_h$ from the
4-lepton data in the high-mass range 620$\div$740 GeV. The result
$(m_h)^{\rm fit} \sim (125 \pm 13)$ GeV reproduces the direct
measurement of the Higgs particle mass and thus supports the idea
that $m_h$ and $M_H$ are the masses of two different excitations of
the same field. Therefore, if we combine with the excess at 680 GeV
in the ATLAS $\gamma\gamma$ distribution, there are now two signals
for a new resonance in the same mass region. Even though,
quantitatively, the global statistical significance of each effect
is modest, still the sharp correlation $\gamma_H-\sigma_R$ in the
4-lepton channel should induce to consider seriously these indications.

\end{abstract}

\ccode{PACS numbers: 11.30.Qc; 12.15.-y; 13.85.-t}


\section{Introduction}

At present, the excitation spectrum of the Higgs field is described
in terms of a single narrow resonance of mass $m_h=$ 125 GeV
associated with the quadratic shape of the effective potential at
its minimum. In a description of Spontaneous Symmetry Breaking (SSB)
as a second-order phase transition, this point of view is well
summarized in the review of the Particle Data Group
\cite{Tanabashi:2018oca} where the scalar potential is expressed as
\BE \label{VPDG} V_{\rm PDG}(\varphi)=-\frac{ 1}{2} m^2_{\rm PDG}
\varphi^2 + \frac{ 1}{4}\lambda_{\rm PDG}\varphi^4 \EE By fixing
$m_{\rm PDG}\sim$ 88.8 GeV and $\lambda_{\rm PDG}\sim 0.13$, this
has a minimum at $|\varphi|=\langle \Phi \rangle\sim$ 246 GeV and a
second derivative $V''_{\rm PDG}(\langle \Phi \rangle)\equiv m^2_h=$
(125 GeV)$^2$.

However, recent lattice simulations of $\Phi^4$ in 4D
\cite{lundow2009critical,Lundow:2010en,akiyama2019phase} support
instead the view of SSB as a (weak) first-order phase transition.
While in the presence of gauge bosons SSB is often described as a
first-order transition, recovering this result in pure $\Phi^4$
requires to replace standard perturbation theory with some
alternative scheme. The implications of a first-order scenario in
pure $\Phi^4$ have not been fully exploited because, with a finite
but very large cutoff, besides the 125 GeV resonance, there could be
another much larger mass scale $M_H$. Since vacuum stability would
depend on $M_H$, SSB could originate within the pure scalar sector
regardless of the other parameters of the theory, e.g. the vector
boson and top quark mass.

To recall how this comes out, we will first summarize the results of
refs.\cite{Consoli:2020nwb,symmetry,memorial} where, as a definite
scheme in which $\Phi^4$ exhibits a (weak) first-order transition,
one explored the original Coleman-Weinberg \cite{Coleman:1973jx}
one-loop calculation and the Gaussian effective potential
\cite{Barnes:1978cd,Stevenson:1985zy}. Indeed, in both cases, SSB
takes place when the quanta of the symmetric phase have a very small
but still positive mass squared. These two calculations,
corresponding to different re-summations of graphs, support each
other and admit the same non-perturbative interpretation: an
effective potential $V_{\rm eff}(\varphi )$ given by some classical
background + zero-point energy of a particle with some
$\varphi-$dependent mass $M(\varphi)$. The peculiarity is that, in
both approximations, by defining $m^2_h$ as $V''_{\rm eff}(\varphi
)$ at the minimum, and $M_H$ as the value of $M(\varphi)$ at the
minimum, one finds the following trend in terms of the ultraviolet
cutoff $\Lambda_s$ \BE \label{scale}  L= \ln (\Lambda_s/M_H) \sim
\frac{1 }{\lambda} ~~ ~~~~~~~~ ~~~~~~~~ ~~~~~~~~ ~~~~~~~~    M^2_H
\sim L m^2_h \gg m^2_h\EE Thus there are two possible
renormalization patterns. A first pattern a) where $M_H$ is cutoff
independent, the effective potential has a finite depth $|{\cal
E}|\sim M^4_H$ and a quadratic shape which vanishes, in units of
$M^2_H$, when $\Lambda_s \to \infty$. A second pattern b) where now
$m_h$ is $\Lambda_s-$ independent and one has the opposite view of a
potential with finite curvature at the minimum but an infinite
depth. With pattern a), the relations $m^2_h=\lambda \langle \Phi
\rangle^2 /3$ and $\lambda\sim 16\pi^2/(3L)$, yielding
cutoff-independent $M_H$ and $\langle \Phi \rangle$, produce the
usual non-interacting continuum limit for the fluctuations around
the minimum of the potential. However, differently from the
2nd-order scenario, the symmetry-restoring temperature $T_c \sim
|{\cal E}|^{1/4}\sim M_H$ is now finite in units of $\langle \Phi
\rangle$. This finiteness can be intuitively explained in terms of
an increasing density $\rho \sim \sqrt{L}$ of $\langle \Phi
\rangle=0$ quanta \cite{Consoli:1999ni} which Bose condense in the
${\bf p}=0$ state and are hidden in the vacuum structure. Therefore
$M^3_H\sim \rho \sqrt{\lambda} $ remains non-zero when $\Lambda_s
\to \infty$ \footnote{This somehow resembles superconductivity where
the energy gap and the critical temperature depend on a collective
coupling $G=g{\cal N}_F$ obtained after re-scaling the tiny two-body
strength $g$, of the electrons in a Cooper pair, by the large
density of states ${\cal N}_F$ at the Fermi surface.}.

To further clarify the $m_h-M_H$ difference, let us recall that the
derivatives of the effective potential produce (minus) the n-point
functions at zero external momentum. Hence $m^2_h$, which is
$V''_{\rm eff}(\varphi )$ at the minimum, is directly the 2-point,
self-energy function $|\Pi(p=0)|$. On the other hand, the zero-point
energy is (one-half of) the trace of the logarithm of the inverse
propagator $G^{-1}(p)=(p^2-\Pi(p))$. Then, after subtracting
constant terms and quadratic divergences, matching the 1-loop
zero-point energy(``$zpe$'') at the minimum gives the relation
\BE\label{general} zpe\sim -\frac{1}{4} \int^{p_{\rm max}}_{p_{\rm
min}} {{ d^4 p}\over{(2\pi)^4}} \frac{\Pi^2(p)}{p^4} \sim-\frac{
\langle \Pi^2(p)\rangle }{64\pi^2} \ln\frac{p^2_{\rm max}}{p^2_{\rm
min}}\sim -\frac{M^4_H}{64\pi^2} \ln\frac{\Lambda^2_s }{M^2_H} \EE
This shows that $M^2_H$ effectively refers to some average value
$|\langle \Pi(p)\rangle| $ at larger $p^2$. A non-trivial momentum
dependence of $\Pi(p)$ would then indicate the coexistence, in the
cutoff theory \footnote{This gives one more argument for the
different cutoff dependence of $m_h$ and $M_H$. Indeed, it is
crucial not to run in contradiction with the ``triviality'' of
$\Phi^4$ which requires a continuum limit with a Gaussian set of
Green's functions and a massive free-field propagator. With this
constraint, for a consistent cutoff theory, there are only two
possibilities when $\Lambda_s \to \infty$: either the usual
perturbative limit $m_h/M_H=1 + O(\lambda) \to 1$, or a non-uniform
scaling of the two masses, see \cite{memorial}.}, of two kinds of
``quasi-particles'', with masses $m_h$ and $M_H$, thus closely
resembling the two branches (phonons and rotons) in the energy
spectrum of superfluid He-4 which is usually considered the
non-relativistic analog of the broken phase.

The existence of a two-mass structure in the cutoff theory was
checked with lattice simulations of the scalar propagator
\cite{Consoli:2020nwb}. Then, by computing $m^2_h$ from the $p\to 0$
limit of $G(p)$ and $M^2_H$ from its behaviour at higher $p^2$, the
lattice data are consistent with a transition between two different
regimes. By analogy with superfluid He-4, where the observed energy
spectrum arises by combining the two quasi-particle spectra of
phonons and rotons, the lattice data were well described in the full
momentum region by the model form \cite{memorial}
\BE
\label{interpol} G(p) \sim \frac{1 - I(p)}{2}\frac{1}{p^2 + m^2_h
}+\frac{1 + I(p)}{2}\frac{1}{p^2 + M^2_h }
\EE
with an interpolating function $I(p)$ which depends on an intermediate momentum scale
$p_0$ and tends to $+1$ for large $p^2\gg p^2_0$ and to $-1$ when
$p^2 \to 0$. Most notably, the lattice data were also consistent
with the expected increasing logarithmic trend $M^2_H\sim L m^2_h$
when approaching the continuum limit \footnote{Note that
Eq.(\ref{interpol}) closely resembles van der Bij's two-pole
propagator \cite{jochum2018} deduced on the basis of very different
arguments, quite unrelated to the effective potential and/or lattice
simulations. If the Higgs field propagator has really a two-pole
structure, radiative corrections will then be sensitive to an
effective mass $m_{\rm eff}$ in the range $m_h \leq m_{\rm eff}\leq
M_H$, see \cite{memorial}. Therefore, it becomes important to
understand how well the mass parameter obtained indirectly from
radiative corrections agrees with the $m_h=$ 125 GeV, measured
directly at LHC. Here we just recall that, for a careful check, it
is essential to take into account the positive $m_{\rm
eff}-\alpha_s(M_z)$ correlation \cite{hioki1,hioki2} where the
relevant $\alpha_s(M_z)$ is the one entering the strong-interaction
correction to the quark-parton model in $\sigma(e+e^- \to hadrons)$
at center of mass energy $Q=M_z$. Since the most complete analysis
of $e+e^- \to hadrons$ data \cite{schmitt}, in the range 20 GeV$\leq
Q\leq$ 209 GeV, indicates an overall 4-sigma excess with a value
$\alpha_s(M_z)\gtrsim $ 0.128, the present view, that the Higgs mass
parameter extracted indirectly from radiative corrections agrees
perfectly with the $m_h=$ 125 GeV measured directly at LHC, is not
free of ambiguities.}.

Since, differently from $m_h$, the larger $M_H$ would remain finite
in units of the weak scale $\langle \Phi \rangle\sim$ 246.2 GeV for
an infinite ultraviolet cutoff, one can derive their proportionality
relation. To this end, let us express $M^2_H$ in terms of $m^2_h L$
through some constant $c_2$, say \BE \label{c2} M^2_H = m^2_h L
\cdot(c_2)^{-1} \EE and replace the leading-order estimate
$\lambda\sim 16\pi^2/(3L)$ in the relation $\lambda= 3 m^2_h/\langle
\Phi \rangle^2$. Then $M_H$ and $\langle \Phi \rangle$ are related
through a cutoff-independent constant $K$ \BE M_H=K \langle \Phi
\rangle\EE with $K \sim (4\pi/3)\cdot (c_2)^{-1/2}$. Since, from a
fit to the lattice propagator \cite{Consoli:2020nwb}, we found
$(c_2)^{-1/2} = 0.67 \pm 0.01 ~({\rm stat}) \pm 0.02 ~({\rm sys})$
this gives the estimate \BE \label{prediction} M_H = 690 \pm 10
~({\rm stat}) \pm 20 ~({\rm sys})~ {\rm GeV}\EE After having
summarized the main theoretical framework of
\cite{Consoli:2020nwb,symmetry,memorial}, we will first describe in
Sect.2 the expected phenomenology of the new resonance. We will then
compare in Sect.3 with the ATLAS 4-lepton data \cite{ATLAS2} which
indicate an excess of events  in the same mass region of
Eq.(\ref{prediction}). Even more significantly, in our picture there
is a particular correlation with the lower-resonance mass at 125 GeV
which is reproduced to high accuracy by the ATLAS data. Finally,
Sect.4 will contain a summary, a discussion of the presently
available CMS 4-lepton events and our conclusions, also on the basis
of a (local) 3-sigma excess at 680 GeV observed in the ATLAS
$\gamma\gamma$ distribution.

\section{Basic phenomenology of the new resonance}

By accepting the ``triviality'' of $\Phi^4$ theories in 4D, the
$\Lambda_s-$independent combination $3M^2_H/ \langle \Phi \rangle^2=
3K^2$ {\it cannot} represent a coupling entering observable
processes. Instead, as anticipated, from the relation $|{\cal
E}|\sim M^4_H$, it would be natural to consider $3K^2$ as a
collective self-interaction of the vacuum condensate
\cite{Consoli:1999ni} whose effects are fully re-absorbed into the
vacuum structure. In this sense, the constant $3K^2$ is basically
different from the coupling $\lambda$ governed by the
$\beta-$function \BE \label{beta} \ln\frac{\mu}{\Lambda_s}=
\int^\lambda_{\lambda_0}~\frac{dx}{\beta(x)} \EE For  $\beta(x) =
3x^2/(16\pi^2) + O(x^3)$, whatever the contact coupling $\lambda_0$
at the asymptotically large $\Lambda_s$, at finite scales $\mu\sim
M_H$ this gives $\lambda\sim 16\pi^2/(3L)$ with $L= \ln
(\Lambda_s/M_H)$.

As emphasized in \cite{Castorina:2007ng,memorial}, there is no
contradiction with the original calculation \cite{lee} in the
unitary gauge. This could give the impression that, with a mass
$M_H$ in the scalar propagator, very high-energy $W_LW_L$ scattering
is indeed similar to $\chi\chi$ Goldstone boson scattering with a
contact coupling $\lambda_0= 3K^2$. However, this is just the result
of a tree approximation with the same coupling at all momentum
scale. To find the $W_LW_L$ scattering amplitude at some scale $\mu$
one should first use the $\beta-$function to re-sum the higher-order
effects in $\chi\chi$ scattering \BE A(\chi\chi \to \chi
\chi)\Big|_{g_{\rm gauge}=0}\sim \lambda \sim \frac{1 }{\ln
(\Lambda_s/\mu)}\EE and then use the Equivalence Theorem
\cite{Cornwall:1974km,Chanowitz:1985hj,Bagger:1989fc} which gives
\BE A(W_LW_L \to W_LW_L)= [1 +O(g^2_{\rm gauge})]~A(\chi\chi \to
\chi \chi)\Big|_{g_{\rm gauge}=0}=O(\lambda)\EE Thus the large
coupling $\lambda_0=3K^2$ is actually replaced by the much smaller
coupling \BE \lambda= \frac{3m^2_h }{\langle \Phi \rangle^2 }= 3K^2
~\frac{m^2_h }{M^2_H}\sim 1/L \EE For the same reason, the
conventional large width into longitudinal vector bosons computed
with $\lambda_0= 3K^2$, say $\Gamma^{\rm conv}(H \to W_LW_L) \sim
M^3_H/\langle \Phi \rangle^2$, should instead be rescaled by
$(\lambda/3K^2)=m^2_h/M^2_H$. This gives \BE \label{aux} \Gamma(H\to
W_LW_L) \sim \frac{m^2_h }{M^2_H} ~\Gamma^{\rm conv}(M_H \to W_LW_L)
\sim M_H ~\frac{m^2_h }{\langle \Phi \rangle^2} \EE where $M_H$
indicates the available phase space in the decay and ${m^2_h
}/{\langle \Phi \rangle^2} $ the interaction strength. If the
heavier state couples to longitudinal W's with the same typical
strength of the low-mass state it would represent a relatively
narrow resonance.

With these premises, it was proposed \cite{cc2020,memorial} that
this hypothetical new resonance could naturally fit with some
localized excess of 4-lepton events in the ATLAS data around 680 GeV
\cite{ATLAS2}. Of course, the 4-lepton channel is just one possible
decay channel and, for a complete analysis, one should also look at
the other final states.  However, this final state fixes the
kinematics in an incomparable way and, for this reason, is
considered the ``golden'' channel to exploit the existence of a
heavy Higgs resonance. At the same time, the bulk of the effect can
be analyzed at an elementary level. Therefore it is natural to start
from here.

The main new aspect is the strong reduction of the conventional
width in Eq.(\ref{aux}). By assuming for definiteness the reference
value $M_H=$ 700 GeV, where $ \Gamma^{\rm conv}( H \to ZZ)\sim
56.7~{\rm GeV}$ \cite{Djouadi:2005gi,handbook}, this gives \BE
\label{width18} \Gamma( H \to ZZ)\sim \frac{m^2_h}{(700~{\rm
GeV})^2}~56.7~{\rm GeV} \EE so that for $m_h=125$ GeV one finds
$\Gamma( H \to ZZ) \sim$ 1.8~{\rm GeV}.

With this premise, in \cite{cc2020,memorial} one was also assuming
from refs.\cite{Djouadi:2005gi,handbook} the values $\Gamma(H\to
{\rm fermions+ gluons+ photons...})\sim 28~{\rm GeV}$ and the ratio
$ \Gamma( H \to W^+W^-)/\Gamma( H \to ZZ)\sim$ 2.03 deducing a total
width $\Gamma( H \to all) \sim 33.5~{\rm GeV}$ and a fraction $ B( H
\to ZZ)\sim (1.8~/~33.5) \sim 0.054$. However, these two estimates
were not taking into account the new, additional contributions to
the total width due to the decays of the heavier state into the
lower-mass state at 125 GeV. These include the two-body process
$H\to hh$, the three-body processes $H\to hhh$, $H\to hZZ$, $H\to
hW^+W^-$ and all higher-multiplicity final states allowed by phase
space. For this reason, the above value $\Gamma( H \to all) \sim
33.5~{\rm GeV}$ should only be considered as a lower bound. For the
same reason, the fraction $ B( H \to ZZ)\sim (1.8~/~33.5) \sim
0.054$ should also be considered as an upper bound.

Since it is not easy to evaluate these additional contributions,
here, to compare with the ATLAS data, we will perform a test of our
picture that does {\it not} require the knowledge of the total width
but just relies on two assumptions:

~~ a) a resonant 4-lepton production by the hypothetical heavy $H$
which proceeds through the process $H \to ZZ \to 4l$

~~ b) the estimate in Eq.(\ref{width18}) together with the linear
scaling law of our model for small variations around $M_H=700$
GeV\BE \label{rel1} \Gamma( H \to ZZ)\sim \frac{M_H}{700~{\rm
GeV}}\cdot\frac{m^2_h}{(700~{\rm GeV})^2}~56.7~{\rm GeV} \EE
Therefore, by defining $\gamma_H=\Gamma( H \to all)/M_H$, we find a
fraction \BE B( H \to ZZ)= \frac {\Gamma( H \to ZZ)}{\Gamma( H \to
all)}\sim \frac {1}{\gamma_H}\cdot \frac{56.7}{700}
\cdot\frac{m^2_h}{(700~{\rm GeV})^2} \EE that will be replaced in
the cross section approximated by on-shell branching ratios
\BE\label{exp3} \sigma_R (pp\to H\to 4l)\sim \sigma (pp\to H)\cdot
B( H \to ZZ) \cdot 4 B^2(Z \to l^+l^-) \EE This should be a good
approximation for a relatively narrow resonance, where the effects
of its virtuality should be small, so that one gets the anticipated
correlation \BE\label{exp33} \gamma_H\cdot \sigma_R (pp\to H\to
4l)\sim \sigma (pp\to H)\cdot \frac{56.7}{700}\cdot
\frac{m^2_h}{(700~{\rm GeV})^2}\cdot 4 B^2(Z \to l^+l^-) \EE Since
$4B^2( Z \to l^+l^-)\sim 0.0045$, the last ingredient we need is the
total production cross section $\sigma (pp\to H)$. As discussed in
\cite{cc2020,memorial}, the relevant production mechanism in our
picture is through the gluon-gluon Fusion (ggF) process. In fact,
the other production through Vector-Boson Fusion (VBF) plays no
role. The point is that the $VV\to H$ process (here $VV=W^+W^-, ZZ$)
is the inverse of the $H\to VV$ decay so that $\sigma^{\rm
VBF}(pp\to H)$ can be expressed \cite{kane} as a convolution with
the parton densities of the same Higgs resonance decay width. The
importance given traditionally to this mechanism depends crucially
on the conventional large width into longitudinal $W$'s and $Z$'s
computed with the $3K^2$ coupling. In our case, where this width is
rescaled by the small ratio $(125/700)^2\sim 0.032$, one finds
$\sigma^{\rm VBF}(pp\to H)~ \lesssim 10~ $ fb which can be safely
neglected.
\begin{table}[htb]
\tbl{We report the ggF cross section in fb to produce a
heavy Higgs resonance at $\sqrt{s}=$ 8 and 13 TeV. The ratios of the
two cross sections, respectively 4.311, 4.393 and 4.477, for $M_H=$
660, 680 and 700 GeV, and the 8 TeV values were taken from the
updated Handbook of Higgs cross sections in the CERN yellow report
\cite{yellow}. No theoretical uncertainty is reported.}
{\begin{tabular}{@{}ccc@{}} \toprule
$M_H$[GeV] &~ $\sigma_{\rm gg}$(8 TeV)& $\sigma_{\rm gg}$(13 TeV)  \\
\hline
660 & 315.3 & 1359.26   \\
\hline
680 & 268.2 & 1178.20  \\
\hline
700& 229.0 & 1025.23 \\
\botrule
\end{tabular}}
\end{table}

Thus, we will replace $\sigma (pp\to H)\to \sigma^{\rm ggF} (pp\to
H)$ in Eq.(\ref{exp33}) and use the ggF cross sections taken from
the updated Handbook of Higgs cross sections \cite{yellow} and
reported in Table 1. For 13 TeV pp collisions, and taking into
account a typical $\pm 15\%$ uncertainty (due to the choice of the
parton distributions, of the QCD scale and other effects), we will
adopt here the estimate $\sigma^{\rm ggF} (pp\to H)= $ 1180(180) fb
which, in our case, also accounts for the range $M_H=660\div 700$
GeV. Therefore, by fixing $m_h=$ 125 GeV, we arrive to a theoretical
prediction which, for not too large $\gamma_H$ where Eq.(\ref{exp3})
becomes inadequate, is formally insensitive to the value of the
total width and can be compared with the ATLAS data \BE\label{exp34}
[\gamma_H\cdot \sigma_R (pp\to H\to 4l)]^{\rm theor} \sim (0.0137
\pm 0.0021)~ {\rm fb} \EE

\section{Analysis of the ATLAS 4-lepton  events}

To check the precise correlation in Eq.(\ref{exp34}), we have
considered the full ATLAS sample \cite{ATLAS2} of 4-lepton data for
luminosity 139 fb$^{-1}$ and in the region of invariant mass
$M_{4l}=620\div740$ GeV ($l=e,\mu$) which extends about $\pm $ 60
GeV around our mass value  $M_H = 690 \pm 10 ~({\rm stat}) \pm 20
~({\rm sys})~ {\rm GeV}$.

Now, Eq.(\ref{exp34}) accounts only for production through the ggF
mechanism and ignores the VBF-production mode which plays no role in
our picture. Therefore, we should compare with that subset of data
that, for their typical characteristics, admit this interpretation.
To this end, the ATLAS experiment has performed a multivariate
analysis (MVA) of the ggF production mode which combines a
multilayer perceptron (MLP) and one or two recurrent neural networks
(rNN). The outputs of the MLP and rNN(s) are concatenated so as to
produce an event score. In this way, depending on the score, the ggF
events are divided into four mutually exclusive categories:
ggF-MVA-high-4$\mu$, ggF-MVA-high- 2e2$\mu$, ggF-MVA-high-4e,
ggF-MVA-low. The four sets of events were extracted from the
corresponding HEPData file \cite{atlas4lHEPData} and are reported in
Table 2.
\begin{table}[htb]
\tbl{At the various 4-lepton invariant mass $M_{4l} \equiv E$, we
report the ATLAS events for the four different categories of the ggF
production mode and their total number.}
{\begin{tabular}{@{}cccccc@{}} \toprule
$M_{4l}$[GeV] &~MVA-high-4$\mu$ & MVA-high-2e2$\mu$& MVA-high-4e& MVA-low&ToT  \\
\hline
635(15) & 2& 0 & 1 & 7&10   \\
\hline
665(15) &0&2&2&17 &21  \\
\hline
695(15)& 1&0&1&9&11 \\
\hline
725(15) & 0& 1& 0& 3& 4 \\
\botrule
\end{tabular}}
\end{table}

By defining $M_{4l}= E$ and $s=E^2$ we have then transformed the
total number of the ggF 4-lepton events in Table 2 into cross
sections for the given luminosity 139 fb$^{-1}$. As in
refs.\cite{cc2020,memorial}, we then assumed the interference of a
resonating amplitude $A^{R}(s)\sim 1/(s - M^2_R)$ with a slowly
varying background $A^{B}(s)$. For a positive interference below
peak, setting $M^2_R=M^2_H -i M_H \Gamma_H$, this gives a total
cross section \BE \label{sigmat} \sigma_T=\sigma_B
-\frac{2(s-M^2_H)~\Gamma_H M_H}{(s-M^2_H)^2+ (\Gamma_H
M_H)^2}~\sqrt{\sigma_B\sigma_R} +\frac{(\Gamma_H M_H)^2
}{(s-M^2_H)^2+ (\Gamma_H M_H)^2}~ \sigma_R\EE  where, in principle,
both the average background $\sigma_B$, at the central energy 680
GeV, and the resonating peak cross-section $\sigma_R$ can be treated
as free parameters.

\begin{table}[htb]
\tbl{For each $\gamma_H$ we report the values of $M_H$, the
resonating cross section $\sigma_R$  and the corresponding product
$k=\gamma_H\cdot\sigma_R$ which are obtained from a fit with
Eq.(\ref{sigmat}) to the total number of ATLAS events in Table 2.}
{\begin{tabular}{@{}cccc@{}} \toprule
$\gamma_H$ & $M_H$ [GeV]& $\sigma_R$ [fb] & $k=\gamma_H\cdot\sigma_R$ [fb] \\
\hline
0.05  &678(6) & 0.218(39)&0.0109(20)  \\
\hline
0.06  &676(7) & 0.191(30)&0.0115(18)  \\
\hline
0.07  &673(10)& 0.174(26)& 0.0122(18)\\
\hline
0.08  &669(20)& 0.161(24)&0.0129(19) \\
\hline
0.09  &668(16)& 0.151(22)&0.0136(20) \\
\hline
0.10  &668(15)& 0.141(21)&0.0141(21)  \\
\hline
0.11  &669(15)& 0.133(21)&0.0146(23)  \\
\hline
0.12  &670(16)& 0.125(22)&0.0150(26) \\
\hline
0.13  &672(17)& 0.118(23)&0.0153(30)  \\
\hline
0.14  &673(19)& 0.112(26)&0.0157(36) \\
\hline
0.15  &674(20)& 0.106(29)&0.0159(43) \\
\botrule
\end{tabular}}
\end{table}

\begin{figure}[ht]
\centering
\includegraphics[width=0.5\textwidth,clip]{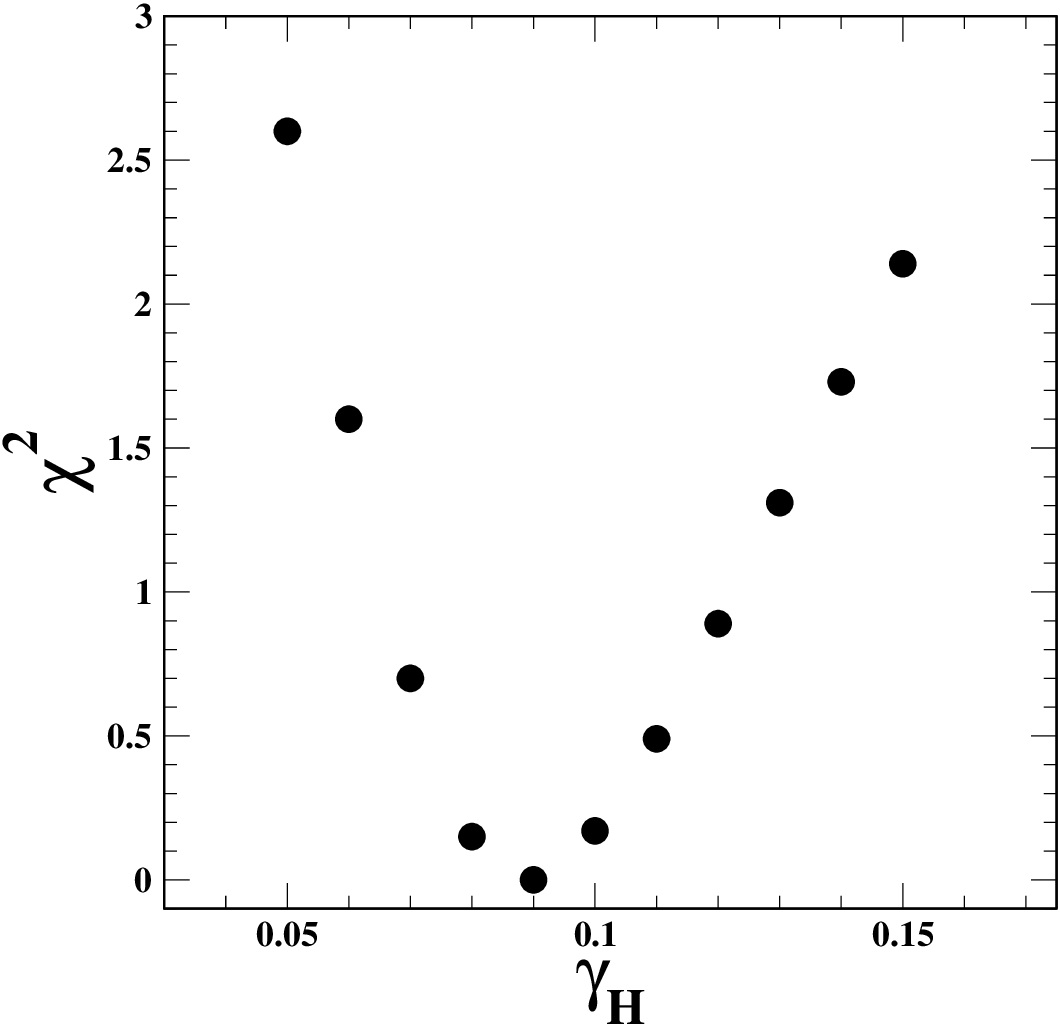}
\caption{{\it At the various values of $\gamma_H$, we report the
chi-square of the fit with Eq.(\ref{sigmat}) to the ATLAS data.}}
\label{chquadro}
\end{figure}

\begin{figure}[ht]
\centering
\includegraphics[width=0.5\textwidth,clip]{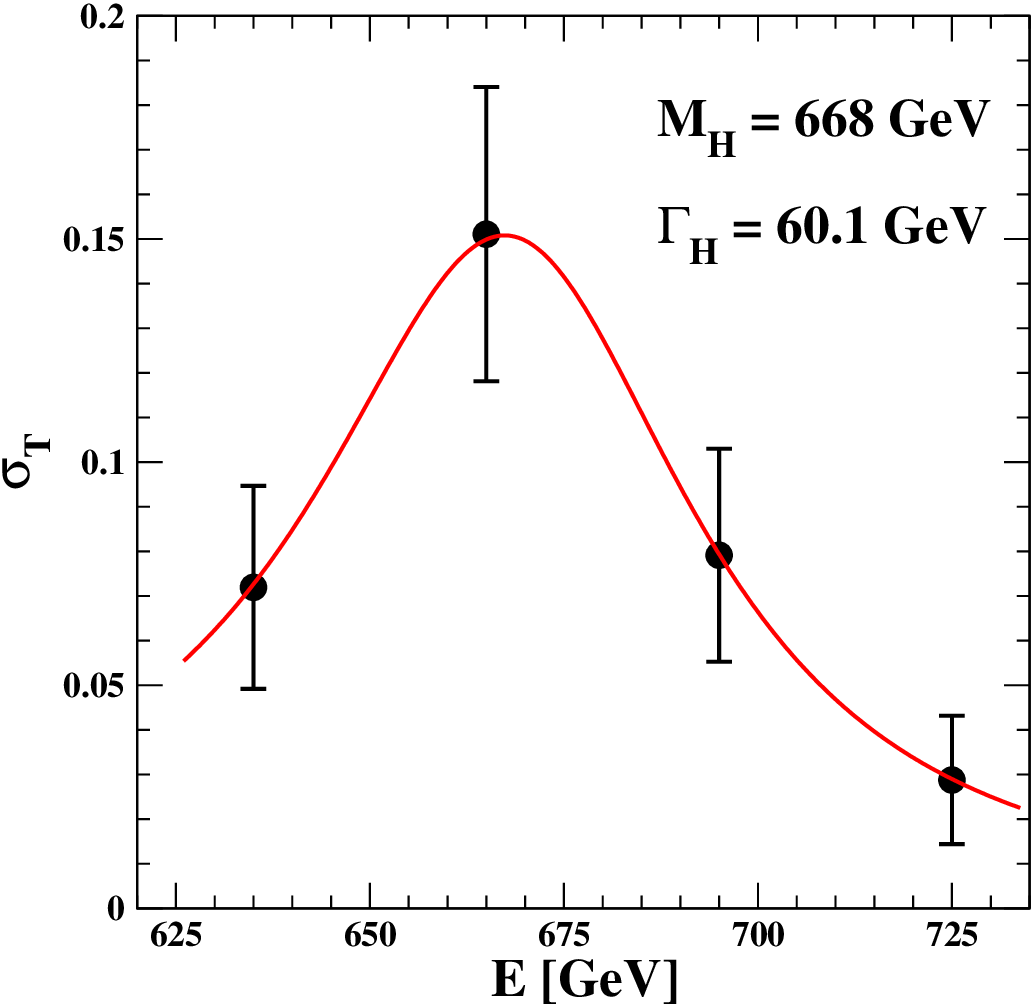}
\caption{ {\it For $\gamma_H=0.09$, we show the fit with
Eq.(\ref{sigmat}) to the ATLAS cross sections in fb.}}
\label{fitatlas}
\end{figure}

\begin{figure}[ht]
\centering
\includegraphics[width=0.5\textwidth,clip]{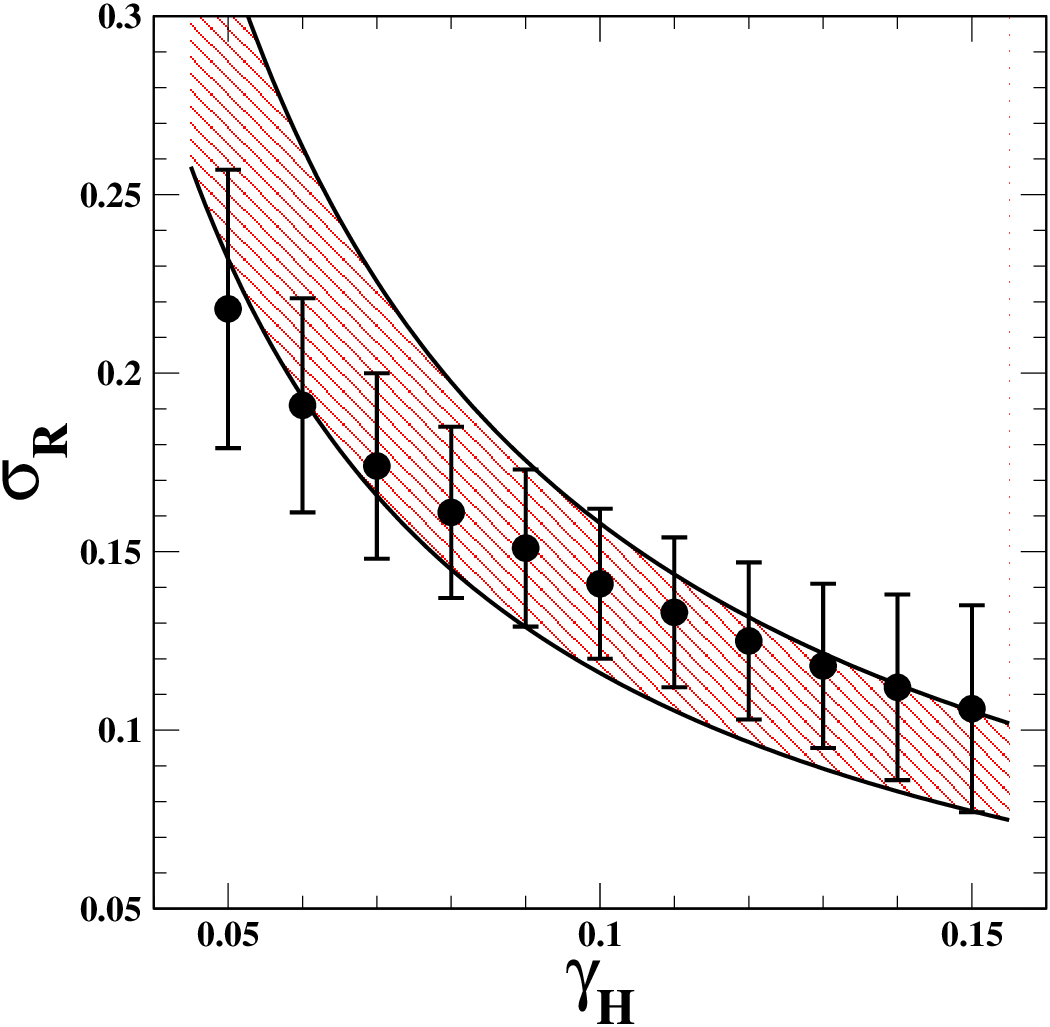}
\caption{ {\it The $\sigma_R$'s of Table 3 are compared with our
theoretical prediction Eq.(\ref{exp34}) represented by the shaded
area enclosed by the two hyperbolae $\sigma_R=(0.0137 \pm
0.0021)/\gamma_H$.} }
\label{correl}
\end{figure}

In a first series of fits to the ATLAS data, for each given
$\gamma_H= \Gamma_H/M_H$, there were 3 free parameters, namely
$M_H$, $\sigma_R$ and $\sigma_B$. As a control, to check the
stability of the results, we then repeated the analysis by assuming
the background to be a decreasing function of energy. To this end,
for each given $\gamma_H$, we considered the central value $\langle
\sigma_B\rangle$ from the first series of fits and replaced the
constant background with a function depending on a slope parameter
$\Delta \sigma_B \ge 0$ \BE \sigma_B(E)= \langle \sigma_B\rangle -
\Delta \sigma_B \cdot \frac{(E-680) }{680}\EE In this second series
of fits, $\Delta \sigma_B$ was further constrained by $0\leq \Delta
\sigma_B \leq \langle \sigma_B\rangle (680/45)$ imposing the
positivity of $\sigma_B(E)$ at the upper limit of the energy range.
Therefore, again, for each given $\gamma_H$, there were 3 free
parameters: $M_H$, $\sigma_R$, and $\Delta \sigma_B $.

The second series of fits did not show any appreciable evidence for
an energy-decreasing background so that we reported in Table 3 the
results obtained with a constant average background. The profile of
the $\chi^2$ as function of $\gamma_H$ and the fit to the ATLAS
cross sections for $\gamma_H=0.09$ are reported respectively in
Fig.1 and in Fig.2 \footnote{The {\it fitted} average background
shows some mild dependence on the input $\gamma_H$ value. In all
cases, however, we found $\langle \sigma_B\rangle^{\rm fit} \lesssim
0.03$ fb, thus indicating an average total background events
$\langle N_B \rangle^{\rm fit} \lesssim$ 17. This is about twice as
small as the background {\it estimated} by ATLAS  $\langle N_B
\rangle^{\rm estimated} \sim$ 36. As a partial explanation for this
difference, we observe that the two external bins at 635(15) and
725(15) GeV of \cite{ATLAS2}, which are less sensitive to the
presence of a resonance, have less events than expected.}.

Finally, to show the very good consistency with our theoretical
prediction Eq.(\ref{exp34}), we have reported in Fig.3 the peak
cross sections of Table 3 and compared with the shaded area enclosed
by the two hyperbolae $\sigma_R=(0.0137 \pm 0.0021)/\gamma_H$. This
picture illustrates how well the observed $\gamma_H-\sigma_R$
correlation in Table 3 is reproduced in our model. In particular,
notice the excellent agreement between Eq.(\ref{exp34}) and the
value $k=\gamma_H\cdot\sigma_R=$0.0136 for $\gamma_H=0.09$ which
gives the minimum $\chi^2$. Finally, a fit to all entries in Table 3
with $\chi^2<1$ gives \BE\label{exp44} [\gamma_H\cdot \sigma_R
(pp\to H\to 4l)]^{\rm fit}=k \sim (0.0137 \pm 0.0008)~ {\rm fb} \EE
This value can then be replaced in the left-hand side of
Eq.(\ref{exp33}) by providing the combined determination \BE
[\sigma(pp\to H)\cdot m^2_h]^{\rm fit}= (1.84 \pm 0.11)\cdot 10^7
~{\rm fb}\cdot {\rm GeV}^2\EE Therefore, with the previous estimate
$\sigma (pp\to H)\sim\sigma^{\rm ggF} (pp\to H)\sim$ 1180(180) fb,
we find \BE \label{mhfit} (m_h)^{\rm fit}\sim (125 \pm 13) ~ {\rm
GeV}\EE whose central value coincides with the measured Higgs
particle mass.

\section{Summary and conclusions}

From the phenomenological analysis of Sect.3, we can draw the
following conclusions:

~~i) by inspection of Table 3, the ATLAS 4-lepton data suggest the
existence of a new resonance $H$ whose mass $M_H=660\div 680$ GeV is
consistent with our prediction Eq.(\ref{prediction}).
Quantitatively, if we look at Fig.2 and compare with the {\it
estimated} background, the local significance of this $M_H$ is about
2.5 $\sigma$ and almost entirely due to the central peak at 665(15)
GeV. However, the global significance, estimated along the lines of
ref.\cite{upcrossing}, is considerably smaller, about 1.4 $\sigma$;

~~ii) by assuming a partial width $\Gamma(H\to ZZ)$ which scales as
in Eq.(\ref{rel1}), we obtain the theoretical prediction
Eq.(\ref{exp34}) which is well consistent with the corresponding
Eq.(\ref{exp44}) obtained from a fit to the ATLAS data in the
high-mass range $620\div740$ GeV. Equivalently, the central value of
the fitted lower-resonance mass $(m_h)^{\rm fit}\sim (125 \pm 13)$
GeV in Eq.(\ref{mhfit}) coincides with the direct, experimental
determination $m_h=$ 125 GeV;

~~iii) consistently with our picture, in the ATLAS analysis there is
no sizeable contribution from the VBF production mode to the new
resonance (on average, only 2 VBF-like events vs. 46 ggF-like
events, see Fig.2e of ref.\cite{ATLAS2});

~~iv) re-obtaining exactly the same central value $m_h=125$ GeV
means that, for $M_H \sim 680$ GeV, a ggF cross section of about
1180 fb and the ATLAS selection criteria of ggF-like events are
consistent to a high degree of precision;

~~v) the correlation successfully reproduced in Fig.3 effectively
eliminates the spin-zero vs. spin-2 ambiguity in the interpretation
of the heavy state.

Therefore, our picture of a second resonance of the Higgs field
finds support in the present ATLAS data. Given the importance of the
issue, we have also attempted a comparison with CMS and looked for
their 4-lepton data in the relevant energy region E=650$\div$700
GeV. Right now, this can only be done with smaller statistical
samples because in the full 137 fb$^{-1}$ CMS analysis
\cite{CMSlarge}, all data in the range 600$\div$800 GeV were
summarized into a single bin of 200 GeV. We have thus compared with
previous reports, for instance the 35.9 fb$^{-1}$ sample shown in
Fig.3 (left) of \cite{CMSsmall}. In spite of its smaller statistics,
this plot is useful because it shows an event distribution strongly
peaked around 660 GeV and which does not slowly decrease with
energy, as expected from the modeled background (on average,there
are 8 events in the range 600$\div$700 GeV and only 1 marginal event
at the very end of the range near 800 GeV).

\begin{figure}[ht]
\centering
\includegraphics[width=0.8\textwidth,clip]{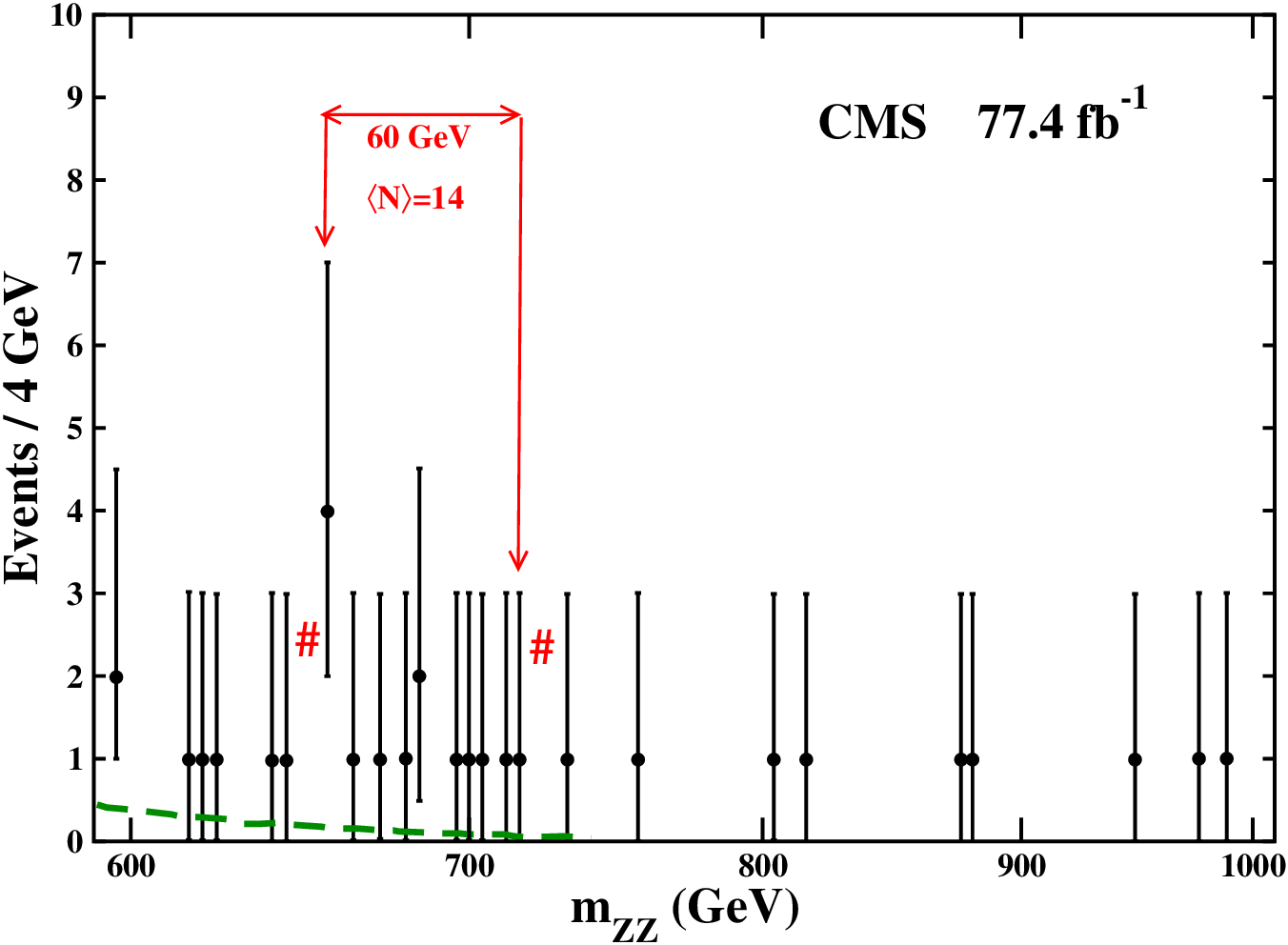}
\caption{ {\it  The 4-lepton data observed by CMS, for integrated
luminosity 77.4 fb$^{-1}$, given in Fig.9 of \cite{CMSreport} as
plotted in an expanded scale by Cea \cite{cea} for the range
$600\div 1000$ GeV. The interval in red color corresponds to
655$\div$715 GeV and solves the problem of overlapping events. } }
\label{ceafigure}
\end{figure}

At present, the largest existing CMS sample which can serve for our
scope refers to integrated luminosity 77.4 fb$^{-1}$ corresponding
to the 2016+2017 data only, see Fig.9 of \cite{CMSreport}. Since the
very compressed scale prevents a straightforward interpretation, we
have taken advantage of Cea's paper \cite{cea} where these CMS data,
in bins of 4 GeV, were extracted and plotted in an expanded scale,
see his Fig.1a) here reported as our Fig.4 \footnote{For the whole
range 600$\div$800 GeV, the 77.4 fb$^{-1}$ sample in Fig.4 gives an
average number of events $\langle {\rm N_{\rm obs}( 4l)}\rangle
=21\div 24$, depending upon the inclusion or not of 3 marginal
events at the extreme left and extreme right of the reported energy
range. After re-scaling by the factor (137/77.4)=1.77, this measured
number corresponds to an extrapolated value $\langle \rm N({\rm
4l})\rangle^{\rm extrapolated}$ = $37\div 42$ which is well
consistent with the actual measurement ${\rm N_{\rm obs}( 4l)} \sim
40\pm 7$ for the full 137 fb$^{-1}$ statistics in
ref.\cite{CMSlarge}. While this shows that the data in Fig.4 form a
consistent subset of the full 137 fb$^{-1}$ sample, still the event
distribution in Fig.4 has not the slowly decreasing trend expected
from the modeled background.}.

To solve the problem of overlapping events, we have then grouped
these data in a single bin of 60 GeV which corresponds approximately
to the range formed by the two central ATLAS bins at E=665(15) and
E=695(15) GeV. On average, there are 14 events that, when scaled by
the luminosity ratio 139/77.4= 1.8, would imply 25 ATLAS events.
This would be in excellent agreement with the ggF-MVA-low category
in Table 2 which give indeed 17+9=26. However, the correspondence
between the two sets of data is still to be clarified and,
hopefully, postponed to a combined analysis of the two
Collaborations.

Finally, we cannot close this paper without mentioning the (local)
3-sigma excess, see Fig.3 of \cite{atlas2gamma}, which is present in
the ATLAS $\gamma\gamma$ distribution for the same invariant-mass
$M_{\gamma\gamma} \sim $ 680 GeV obtained from our analysis of the
4-lepton data. Even though the global statistical significance is
reduced to about 1.5 $\sigma$, by the looking-elsewhere effect
\cite{upcrossing}, still this particular excess of events represents
the highest peak in Fig.3 of \cite{atlas2gamma}. 
Nevertheless, the strong indication for $M_H$ is
the sharp $\gamma_H-\sigma_R$ correlation in the ATLAS 4-lepton
channel. 
\vskip 10 pt
\centerline{\bf Acknowledgments} \vskip 5 pt \noindent M. C. would
like to thank Fabrizio Fabbri for many useful discussions.


\end{document}